\documentclass[]{article}
\pdfoutput=1
\usepackage[utf8]{inputenc}

\usepackage{arxiv}
\usepackage{booktabs} 
\usepackage{listings}
\usepackage{graphicx}
\usepackage{hyperref}
\usepackage{amsmath}
\usepackage{amssymb}

\usepackage[T1]{fontenc}
\usepackage{subcaption} % subfigures
\usepackage{verbatim} % block comments

\author{
Sarah Bird\\
Mozilla\\
\texttt{sbird@mozilla.com}
\And
Vikas Mishra\\
INRIA\\
\texttt{vikas.mishra@inria.fr}
\And
Steven Englehardt\\
Mozilla\\
\texttt{senglehardt@mozilla.com}
\And
Rob Willoughby\\
University of British Colombia\\
\texttt{rwilloug@students.cs.ubc.ca}
\And
David Zeber\\
Mozilla\\
\texttt{dzeber@mozilla.com}
\And
Walter Rudametkin\\
INRIA\\
\texttt{walter.rudametkin@inria.fr}
\And
Martin Lopatka\\
Mozilla\\
\texttt{mlopatka@mozilla.com}
}
   
\title{\huge Actions speak louder than words: Semi-supervised learning for browser fingerprinting detection}
\date{November 1, 2019}

\begin{document}
\maketitle

\begin{abstract}
As online tracking continues to grow, existing anti-tracking and fingerprinting detection techniques that require significant manual input must be augmented. Heuristic approaches to fingerprinting detection are precise but must be carefully curated. Supervised machine learning techniques proposed for detecting tracking require manually generated label-sets. Seeking to overcome these challenges, we present a semi-supervised machine learning approach for detecting fingerprinting scripts. Our approach is based on the core insight that fingerprinting scripts have similar patterns of API access when generating their fingerprints, even though their access patterns may not match exactly. Using this insight, we group scripts by their JavaScript (JS) execution traces and apply a semi-supervised approach to detect new fingerprinting scripts. We detail our methodology and demonstrate its ability to identify the majority of scripts ($\geqslant$94.9\%) identified by existing heuristic techniques. We also show that the approach expands beyond detecting known scripts by surfacing candidate scripts that are likely to include fingerprinting. Through an analysis of these candidate scripts we discovered fingerprinting scripts that were missed by heuristics and for which there are no heuristics. In particular, we identified over one hundred device-class fingerprinting scripts present on hundreds of domains. To the best of our knowledge, this is the first time device-class fingerprinting has been measured in the wild. These successes illustrate the power of a sparse vector representation and semi-supervised learning to complement and extend existing tracking detection techniques.
\end{abstract}


\section{Introduction}
\label{introduction}

Web users have been exposed to a wide variety of privacy intrusions since the introduction of third-party cookies in 1994 \cite{schwartzGivingWebMemory2001}. In the ensuing years there has been a steady increase in the prevalence and sophistication of web tracking to the point where users are tracked by tens of companies on nearly every page they visit \cite{lernerInternetJonesRaiders2016, roesnerDetectingDefendingThirdparty2012, HiddenPerilsCookie, englehardtCookiesThatGive2015, papadopoulosCookieSynchronizationEverything2018, arshadIncludeMeOut2017, bashirTracingInformationFlows2016, bashirDiffusionUserTracking2018}.

In this paper we focus on detecting browser fingerprinting, a tracking technique that involves the collection of device features through active inspection of the device via JavaScript (JS), or through the passive collection of HTTP headers and network identifiers. Fingerprinting is not visible to users and is not covered by browser privacy controls.

Web browsers have recently started to take an active stance in preventing web tracking through stateful (i.e., cookie-based) approaches \cite{wilIntelligentTrackingPrevention2019,campFirefoxNowAvailable}. While stateful tracking techniques are still the most commonly deployed \cite{englehardtOnlineTracking1millionsite2016,fouadTrackingPixelsDetecting2019}, some browsers have chosen not to deploy these restrictions out of fear of an increased use of fingerprinting \cite{BuildingMorePrivate2019}.

Mozilla \cite{SecurityTrackingPolicy} and WebKit \cite{WebKitTrackingPrevention} have released anti-tracking policies that commit to blocking fingerprinting, and several of the major browsers have proposed \cite{lasseyContributeBslasseyPrivacybudget2019} or are experimenting with \cite{campFirefoxNowAvailable,SecurityFingerprintingMozillaWiki} anti-fingerprinting techniques.
Most relevant to this work is Mozilla's content blocking approach, done in partnership with Disconnect \cite{campFirefoxNowAvailable}, which relies on heuristics and manual curation of fingerprinting scripts \cite{MozillaSecurityResearchEnglehardt,MozillaSecurityResearchBird}.

Fingerprinting is difficult to prevent \cite{MitigatingBrowserFingerprinting,vastelFpScannerPrivacyImplications}. Despite a steady rise in the use of fingerprinting by trackers \cite{nikiforakisCookielessMonsterExploring2013, acarFPDetectiveDustingWeb2013, acarWebNeverForgets2014, englehardtOnlineTracking1millionsite2016}, users have few options to protect themselves. While a variety of protection options have been proposed (Section~\ref{sec:anti-fingerprinting-techniques}), the most common way users are protected is through the inclusion of fingerprinting scripts in list-based content blockers. Fingerprinting scripts have historically been added to block lists through ad-hoc, manual curation \cite{DisconnectBlocksNew2014}. Our method can augment this process in a semi-automated way; it detects nearly all fingerprinting scripts flagged by state-of-the-art heuristics while surfacing an expanded set of scripts that can be reviewed by list maintainers. 

Our model employs a semi-supervised approach leveraging the core insight that fingerprinting scripts share similarities in the execution steps they undertake to generate fingerprints.
As such, the problem of detecting fingerprinting scripts aligns with the entity set expansion \cite{zhangliu2011entity} task of identifying items belonging to a class for which we have a limited seed set of examples.
We adopt a methodology analogous to distributional similarity
\cite{pantelcrestanetal2009web, sarmento2007more, zhangliu2011entity}
in a feature space derived from JS execution patterns.

The flexibility of this method enables it to perform well using both lists of fingerprinting scripts generated using state-of-the-art techniques from past research \cite{englehardtOnlineTracking1millionsite2016,dasWebSixthSense2018} as input, as well as lists generated by simple keyword searches.
We apply our methodology to the publicly available OverScripted web crawl dataset \cite{1-OverscriptedDataSet2018} composed of \textasciitilde{}114 million JS calls to 282 APIs from \textasciitilde{}1.3 million unique script URLs, on \textasciitilde{}2.1 million sites.
We show that our methodology detects ($\geqslant$94.9\%) of the fingerprinting scripts detected by state-of-the-art heuristics, while detecting scripts for which there are no current heuristics, as well as hundreds of cases of a device-class fingerprinting.

\textbf{The main contributions} of this work are:
\begin{itemize}
\item
  The representation of JS execution in a vector space model, has not, to the best of our knowledge, been used previously for feature engineering in tracking detection. This representation allows us to compare execution patterns of scripts based on behavioral similarity.
\item
  A flexible semi-supervised approach which is robust to incomplete and small label sets. 
\item
  A strategy for programmatic labeling to ensure that production deployment scenarios aren't constrained by the need to manually generate label sets.
\item 
  A method for surfacing a pool of candidate fingerprinting scripts that are similar to known fingerprinting scripts, but sufficiently different that they are missed by current detection techniques. With this approach we discover device-class fingerprinting in the OverScripted dataset. To the best of our knowledge, this is the first time device-class fingerprinting has been measured in the wild.  
\end{itemize}

\section{Background and related work}
\label{background-related-work}

\subsection{Tracking}
\label{tracking}

Tracking is the gathering and storage of data in order to establish patterns of users' interests, connections, habits, and browsing behavior.
Over the past 20 years, trackers have developed remarkable abilities to gather, store, aggregate, and redistribute such data \cite{lernerInternetJonesRaiders2016}. 
Numerous entities work to connect these digital footprints with real-world identities in a practice known as Identity Resolution \cite{stanhopeStrategicRoleIdentity2016}. 
Data describing users' online activity, including their browsing history, may then be sold to advertisers, marketers, and businesses.

Mathur et al. \cite{mathurCharacterizingUseBrowserBased2018} review users' understanding of and attitude towards tracking.   
They confirm previous studies \cite{urSmartUsefulScary2012,AmericansOpinionsPrivacy2016,mcdonaldAmericansAttitudesInternet2010} finding that most users are uncomfortable with tracking, but they only have a ``basic'' understanding of it.

This supports our motivation to build tools that protect internet users from having their online behavior, in particular their browsing history, collected by companies without transparency and without consent.

Browsing history is most commonly collected using cookies, in particular third-party cookies \cite{acarWebNeverForgets2014,lernerInternetJonesRaiders2016,macbethTrackingTrackersAnalysing2016,englehardtCookiesThatGive2015,englehardtOnlineTracking1millionsite2016,karajWhoTracksMeMonitoring2018}.
Cookie syncing, documented as prevalent in the wild \cite{acarWebNeverForgets2014,englehardtOnlineTracking1millionsite2016,papadopoulosCookieSynchronizationEverything2018}, enhances the value of third-party cookies by providing a mechanism for trackers to exchange datasets about users 
\cite{HiddenPerilsCookie,nguyenPerspectiveFirefoxQuantum,papadopoulosCookieSynchronizationEverything2018}.
Research has shown that trackers have also used a variety of alternative, more invasive tracking techniques, including supercookies \cite{mcdonaldSurveyUseAdobe2011,soltaniFlashCookiesPrivacy2009} and browser fingerprinting \cite{acarWebNeverForgets2014,nikiforakisCookielessMonsterExploring2013,englehardtOnlineTracking1millionsite2016}. 

\subsection{Fingerprinting}
\label{fingerprinting}

Fingerprinting is the collection of device features for the purpose of generating an identifier.
In contrast to cookie-based tracking, fingerprinting does not require the storage of data on the user's device. Instead, device features are collected through active inspection of the device via JavaScript (JS), or through the passive collection of HTTP headers and network identifiers.

Fingerprinting is typically used to generate an identifier that is distinctive enough to identify a browser across visits to different websites. 
Past research disagrees on how identifying a device fingerprint is -- the reported proportion of devices with a distinct fingerprint varies between 84\% \cite{eckersleyHowUniqueYour2010} and 33\% \cite{gomez-boixHidingCrowdAnalysis2018}, depending on the features used and populations sampled. 
Various forms of browser fingerprinting have been described in the literature and documented in the wild since 2009 \cite{acarWebNeverForgets2014,lernerInternetJonesRaiders2016,moweryPixelPerfectFingerprinting2012,eckersleyHowUniqueYour2010,gomez-boixHidingCrowdAnalysis2018,merzdovnikBlockMeIf2017,englehardtOnlineTracking1millionsite2016,acarFPDetectiveDustingWeb2013,starovPrivacyMeterDesigningDeveloping2018,karajWhoTracksMeMonitoring2018,kaizerAutomaticIdentificationJavaScriptoriented2016,laperdrixBeautyBeastDiverting2016,dasWebSixthSense2018,mayerAnyPersonPamphleteer2009}.
Fingerprinting has a variety of uses, including tracking and fraud prevention \cite{dasWebSixthSense2018}.
In our work we focus solely on the \textit{detection} of fingerprinting and do not attempt to discern its purpose.

\subsubsection{Known methods}
\label{known-methods}

There are a variety of known methods of fingerprinting that have been proposed and measured in the wild. 
Alaca \& van Oorschot provide a comprehensive summary \cite{alacaDeviceFingerprintingAugmenting2016} of fingerprinting techniques.
Researchers continue to uncover new techniques, such as the use of mobile device sensors \cite{dasWebSixthSense2018}.
The browser attributes that are known to expose the highest entropy are the list of browser plug-ins, the properties of the HTML Canvas, the User-Agent string, and the list of available fonts \cite{eckersleyHowUniqueYour2010,laperdrixBeautyBeastDiverting2016,gomez-boixHidingCrowdAnalysis2018}. 
For this study, we focus on active fingerprinting techniques -- those that leverage the browser's built-in JS APIs to uncover properties of the device, in particular:

\textbf{Attribute fingerprinting} Originally demonstrated in \cite{mayerAnyPersonPamphleteer2009}, this technique combines characteristics of a device that can be directly queried with JS.
These include most of the attributes exposed through the Navigator and Screen interfaces, such as the device's operating system, screen resolution, and browser version.

\textbf{Canvas fingerprinting} First discovered in 2012 \cite{moweryPixelPerfectFingerprinting2012}, this technique draws content on a hidden HTML canvas element. A device's combination of fonts, rendering engine, and graphics stack all lead to sufficient variation in the rendered image to support fingerprinting. After drawing, the script reads the rendered image as a string, and uses a hash of the string as the identifier.

\textbf{Canvas font fingerprinting} First measured by Englehardt \& Narayanan in 2016 \cite{englehardtOnlineTracking1millionsite2016}, and distinct from canvas fingerprinting, canvas font fingerprinting uses a hidden HTML canvas element to detect the set of fonts installed on a user's device.

\textbf{Audio fingerprinting} First documented in \cite{englehardtOnlineTracking1millionsite2016}, this technique generates and processes an audio signal using the Audio API. The result of this processing can then be hashed and used as part of a fingerprint.

\textbf{WebRTC fingerprinting} The WebRTC API provides a mechanism to discover a user's local IP address \cite{englehardtOnlineTracking1millionsite2016}, that is the IP address assigned to the device by the router, if the device is behind a NAT. The recovered IP address can be used as part of a device fingerprint.

\textbf{Emoji fingerprinting} Laperdrix et al. \cite{laperdrixBeautyBeastDiverting2016} highlight the potential for emojis to be used in fingerprinting since they have their own unicode characters, but highly distinct representations across a range of devices and operating systems.

\textbf{Device-class fingerprinting} In 2016, Bursztein et al. \cite{burszteinPicassoLightweightDevice2016} introduce device-class fingerprinting,  
which seeks not to generate a unique identifier for a given device, but to distinguish classes of devices, even when the device's \texttt{userAgent} or \texttt{navigator} properties are spoofed.

It is important to note that these categories are not mutually exclusive nor are they comprehensive. A single fingerprinting script could use all of the preceding methods or discover new sources of browser entropy \cite{javascriptTemplateSchwarz2017}.

\subsubsection{Prevalence}
\label{prevalence}

In 2013, Acar et al. \cite{acarFPDetectiveDustingWeb2013} were the first to apply a heuristic-based approach to fingerprinting detection, looking for scripts accessing particular combinations of browser APIs. 
Acar et al.'s 2014 paper \cite{acarWebNeverForgets2014} measured the prevalence of Canvas fingerprinting at 5.5\% overall, with more fingerprinting occurring in less popular sites.
Englehardt \& Narayanan obtained more nuanced estimates in their 2016 study \cite{englehardtOnlineTracking1millionsite2016}, which crawled the home pages of the Alexa top 1 million sites.
They found the overall prevalence of Canvas fingerprinting to be 1.6\%, although this number jumped to 5.1\% when restricted to the Alexa top 1,000.
In addition, they were the first to document the prevalence of Canvas font fingerprinting and WebRTC fingerprinting at 0.33\% and 0.07\% overall, and noted that both types were more prevalent among the most popular sites.

\subsection{Anti-fingerprinting techniques}
\label{sec:anti-fingerprinting-techniques}

There are two general approaches to protecting against fingerprinting: limiting the entropy exposed by the set of Web APIs or detecting and blocking specific scripts that participate in fingerprinting. These approaches are not mutually exclusive.

\textbf{Limiting API entropy.} The entropy exposed by the set of Web APIs can be limited by techniques varying from completely disabling JavasScript \cite{gomez-boixHidingCrowdAnalysis2018} to limiting the way each individual API operates, as in done in the Tor Browser \cite{DesignImplementationTor}. Along these lines, Safari \cite{WhatYouNeed}, Brave \cite{NextGenerationBrave2019}, and Firefox \cite{olejnikBatteryStatusNot} have limited or removed certain APIs that have been found to be used by fingerprinting scripts. Most recently, Google has proposed imposing a privacy budget on scripts, which is consumed with each new API access \cite{lasseyContributeBslasseyPrivacybudget2019}. However, these techniques all have the potential to introduce site breakage \cite{mazelComparisonWebPrivacy2017,1507517FingerprintingbreakageMETA}. While the W3C has recommended that new specifications should try to limit the entropy a specification exposes \cite{MitigatingBrowserFingerprinting}, they have also deemed it impractical to fully prevent fingerprinting using these approaches \cite{UnsanctionedWebTracking, MitigatingBrowserFingerprinting}.

\textbf{Detecting and blocking scripts.} Ad-blocking and tracker-blocking lists have long included companies that participate in fingerprinting \cite{DisconnectBlocksNew2014}, and thus the extensions and browsers that use these lists (e.g., AdBlock Plus, Disconnect, Firefox, Ghostery, and uBlock Origin) offer partial protection from fingerprinting scripts. However, research has shown these lists are only partially effective \cite{merzdovnikBlockMeIf2017,macbethTrackingTrackersAnalysing2016,gugelmannAutomatedApproachComplementing2015,ikramSeamlessTrackingFreeWeb2017,englehardtOnlineTracking1millionsite2016}. For example, Merzdovnik et al. \cite{merzdovnikBlockMeIf2017} note that fingerprinting scripts that had been known in the literature for three years had not been added to filter lists. All of these solutions suffer from the core problem that lists require constant maintenance to discover and evaluate potential fingerprinting scripts. By providing a robust method for detecting and surfacing new fingerprinting scripts, our method can lower the cost of such maintenance tasks and improve the quality of these lists.

\subsection{Detection}
\label{sec:detection}

The simplest approach to detect fingerprinting, used by \cite{lernerInternetJonesRaiders2016} and \cite{starovPrivacyMeterDesigningDeveloping2018}, is to gather information on the number of JS calls made to APIs known to be used for fingerprinting.

This can be taken a step further through the development of more sophisticated rule-sets that look for certain APIs to be hit but not others, calls to be made a minimum number of repeated times, or for specific values to be set.
This enables more precise measurement, and offers the ability to distinguish between different types of fingerprinting \cite{acarWebNeverForgets2014,englehardtOnlineTracking1millionsite2016,acarFPDetectiveDustingWeb2013,dasWebSixthSense2018}. 
The Cliqz group \cite{macbethTrackingTrackersAnalysing2016,karajWhoTracksMeMonitoring2018} have access to browsing data from their users and apply rule-sets to look for various features of HTTP traffic in order to identify a range of tracking including fingerprinting.

We call these rule-based approaches \textit{heuristic} approaches, as distinct from statistical approaches that ``learn'' patterns using machine learning techniques. A number of studies have demonstrated using supervised machine learning techniques to identify trackers, block ads, or identify anti-ad blocking activity. Using supervised classifiers, \cite{iqbalAdWarsRetrospective2017} and \cite{ikramSeamlessTrackingFreeWeb2017} parse JS script code and generate features based on its syntax and semantics. Iqbal et al. \cite{iqbalAdGraphMachineLearning2018} use features from  HTTP, HTML, and JS to build an ad-blocking classifier. Shuba et al. \cite{shubaNoMoAdsEffectiveEfficient2018} explore ad-blocking on mobile phones using the packets generated by mobile apps. Kalavri et al. \cite{kalavriPackWolvesCommunity2016} build a graph of co-occurrences of third-party content. They observe communities of trackers and use this graph structure to then successfully predict and label trackers. Gugelmann et al. \cite{gugelmannAutomatedApproachComplementing2015} analyze 24 hours of HTTP traffic from a university network to derive features of the traffic that serve as good predictors for tracking, and use their model to detect new trackers.

Das et al. \cite{dasWebSixthSense2018} use an unsupervised clustering analysis to identify collections of certain types of fingerprinting activity for identified fingerprinting scripts, but this is distinct from using automated techniques to discover fingerprinting or tracking.

One challenge that all these studies face is the lack of ground truth against which results can be compared. Many use block lists for this purpose, such as EasyList \cite{EasyListOverview} for ad blocking or EasyPrivacy \cite{EasyListPolicy} for tracker blocking.
However, these lists are incomplete \cite{merzdovnikBlockMeIf2017,iqbalAdGraphMachineLearning2018}, and do not differentiate between entities which use traditional tracking techniques and those which fingerprint.
Merzdovnik et al. \cite{merzdovnikBlockMeIf2017} use Acar et al. \cite{acarWebNeverForgets2014} and Englehardt \& Narayanan \cite{englehardtOnlineTracking1millionsite2016} as the basis for determining the effectiveness of blocking tools with respect to fingerprinting.
Many studies \cite{gugelmannAutomatedApproachComplementing2015,mugheesDetectingAdblockersWild2017,liTrackAdvisorTakingBack2015,ikramSeamlessTrackingFreeWeb2017} have had to make their own manually labeled datasets to train or evaluate their work.

Manually reviewing scripts for supervised learning is time-consuming and not readily scalable. Ikram et al. \cite{ikramSeamlessTrackingFreeWeb2017} suggest crowd-sourcing could be used, but that still requires resources and time to orchestrate, manage, and monitor for abuse.

Heuristic generation can be very precise, but it has some limitations: it is time-consuming; once a heuristic is known it is subject to
evasion; their publication may increase the dissemination of a technique; and it may not be possible to develop heuristics for less specific forms of fingerprinting, like attribute fingerprinting.
Additionally, subtleties in implementation can lead to significant differences in how many fingerprinting scripts each set of heuristics identify.

The methodology presented in this paper seeks to complement these existing techniques and alleviate some of the challenges associated with these methods. 
We are not aware of any comparable work using a semi-supervised approach to attempt to isolate fingerprinting scripts. 

\section{Methodology}
\label{methodology}

\subsection{Dataset requirements}
\label{sec:data_collection}

Our model is designed to work on crawl datasets that collect JS execution data, such as those gathered by OpenWPM. It uses the following fields:
\begin{itemize}
\item
  \texttt{script\_url} - URL of the script that executed the call (may be empty if JS was inline)
\item
  \texttt{func\_name} - name of the function where the call originated, can be empty for anonymous functions
\item
  \texttt{symbol} - the JS API that the call accessed, e.g., \texttt{window\allowbreak{}.document\allowbreak{}.cookie},  \texttt{HTMLCanvasElement\allowbreak{}.toDataURL}, \texttt{window\allowbreak{}.navigator\allowbreak{}.platform}
\end{itemize}

\subsubsection{URL cleaning}
\label{subsec:get_clean_script}
When we consider crawl data and identifying scripts, we typically are looking for a URL of the form \texttt{duckduckgo.com/a/script.js}. Real URLs however have significant additional information. In particular, parameter strings can be very long. Throughout our work we regularly ``clean'' scripts in order to ensure we consider \texttt{\textbf{https}://b.com/a/script.js\textbf{?q=b1}} and \texttt{\textbf{http}://b.com/a/script.js\textbf{?q=b2}} as the same URL. To do this we strip the URL of the protocol, parameters, and fragments. For example,
\texttt{https://duckduckgo.com/a/script.js?q=easyprivacy+list\&t=ffab\&ia=web}
becomes \texttt{duckduckgo.com/a/script.js}.

\subsection{Snippet preparation}
\label{sec:snippet_preparation}
We define \textit{snippets} as representations of JS execution data in symbol space, grouped by features engineered from the call data properties. 
This grouping step is an important part of our methodology,
as it determines the granularity at which we measure ``behaviour'', and different groupings lead to dramatically different results.
The idea is for snippets to be representative chunks of execution that are comparable over all scripts.
The trade-off lies in finding a grouping that ``bubbles up'' sufficient characteristic patterns to allow for meaningful comparisons, but remains granular enough that distinct clusters can be distinguished.
Table \ref{tbl:input_data} gives an example of a dataset. Notice that in this example data, \texttt{window.document.cookie} was referenced twice in the \texttt{ga.js} script from the same location. This is common in the wild. For example, in the case of canvas font fingerprinting, the same text is rendered multiple times with different fonts and then measured.
This causes one script to yield dozens of calls to \texttt{CanvasRenderingContext2D.measureText}.

\begin{table}[htbp]
\centering
\begin{tabular}{clll}
    \emph{i} & script\_url & symbol & func\_name \\
    \midrule
    0 & ggl.com/ga.js & window.navigator & function1 \\
    1 & ggl.com/ga.js & window.document.cookie & function1 \\
    2 & ggl.com/ga.js & window.document.cookie & function2 \\
    3 & ggl.com/msc.js & window.navigator & functiona \\
    4 & ggl.com/ga.js & window.navigator & function1 \\
    5 & ggl.com/ga.js & window.document.cookie & function1 \\
    6 & ddr.com/bo.js & window.navigator & dance \\
    \bottomrule
    \vspace{0.1cm}
\end{tabular}
\caption{An example of raw call data (many columns omitted)}
\label{tbl:input_data}
\end{table}

The grouping scheme is used to create snippets.
A snippet is the call count for each of the JS symbols executed by that group.
We create a table of snippets,
each row of which represents one of the the snippets and its corresponding vector of counts per JS symbol.

To choose a grouping scheme, we explored \texttt{location} (the URL of the page on which a script was loaded), \texttt{script\_url}, and \texttt{func\_name} fields from the dataset. For intuition on the tradeoffs involved, consider the inclusion, or not, of the \texttt{location} parameter when combined with the Google analytics script \texttt{ga.js}. 
Including the \texttt{location} in the grouping scheme will result in separate snippets for every site that links to \texttt{ga.js}. 
Conversely, if we do not include \texttt{location} we will have only one snippet for \texttt{ga.js}. 
With the latter we are able to more readily discern \emph{different scripts} with \emph{similar execution patterns}, rather than identifying the same script over and over again.

In the end, we build our grouping with the fully qualified domain name of the \texttt{script\_url}, the end path of the \texttt{script\_url}, and the \texttt{func\_name}. 
Given a realistic \texttt{script\_url}, say \texttt{https://www.alaskaair.com/px/client/main.min.js?param1=2}, the process to create a snippet using our chosen grouping scheme is as follows:
\begin{enumerate}
\def\labelenumi{\arabic{enumi}.}
\item
  extract the fully qualified domain of \texttt{script\_url}: \texttt{alaskaair.com};
\item
  concatenate the end of the path of \texttt{script\_url}:
  \texttt{main.min.js};
\item
  concatenate the \texttt{func\_name}; and
\item
  aggregate the calls per JS symbol for this group.
\end{enumerate}
Table \ref{tbl:chosen_scheme} shows an example output of our chosen grouping scheme.

\begin{table}[htbp]
\centering
\begin{tabular}{lcc}
    Snippet & \texttt{window} & \texttt{window} \\
     & \texttt{.navigator} & \texttt{.document.cookie} \\
    \midrule
    ggl.com\textbar{}\textbar{}ga.js\textbar{}\textbar{}function1 & 2 & 2 \\
    ggl.com\textbar{}\textbar{}ga.js\textbar{}\textbar{}function2 & 0 & 1 \\
    ggl.com\textbar{}\textbar{}msc.js\textbar{}\textbar{}functiona & 1 & 0 \\
    ggl.com\textbar{}\textbar{}bo.js\textbar{}\textbar{}function4 & 1 & 0 \\
    \bottomrule
\end{tabular}
\vspace{0.2cm}
\caption{Example of the chosen grouping scheme using the fully qualified domain of \texttt{script\_url}, the end path of the \texttt{script\_url}, and the \texttt{func\_name} before normalization.}
\label{tbl:chosen_scheme}
\end{table}

The use of \texttt{func\_name} proved to be a critical addition for achieving the right level of granularity. 
We take an example from the popular fingerprinting library \texttt{fingerprintjs2} \cite{vasilyevModernFlexibleBrowser2019a}.
Figure~\ref{getCanvasFP} shows a function from this library. 
This function encapsulates a meaningful chunk of code that builds a canvas fingerprint. 
A minified version of this code may not have human readable function or variable names, but the function would have the same JS execution pattern that can be captured by snippet data. 
This is a core insight underpinning our model: code with similar behavior will have a similar JS execution trace. For example, over many implementations of canvas fingerprinting, we expect there to be similar blocks of code that perform highly similar behaviors. Our snippet representation allows us to identify these similar blocks of code and subsequently link them back to the scripts that execute them.

\begin{figure}[htbp]
\centering
\begin{lstlisting}
var getCanvasFp = function (options) {
    var result = []
    var canvas = document.createElement('canvas')
    canvas.width = 2000
    canvas.height = 200
    canvas.style.display = 'inline'
    var ctx = canvas.getContext('2d')
    ctx.rect(0, 0, 10, 10)
    ctx.rect(2, 2, 6, 6)
    ......
}
\end{lstlisting}
\vspace{-0.5cm}
\caption{Code Sample from fingerprintjs2 \cite{vasilyevModernFlexibleBrowser2019a}}
\label{getCanvasFP}
\end{figure}

The final step of snippet preparation is to normalize the call counts of each symbol in the table of snippets in order to help patterns of interest stand out.
In our case, we are interested in comparing different snippets; each snippet is a table row, so a row-wise normalization was applied such that each row of JS symbol execution counts sums to 1.

One attractive feature of this methodology is that new scripts and rows can be added to the constructed score without needing to recompute the normalization over the whole dataset. Additionally, this approach avoids some of the shortcomings of static analysis methods that can be used to compliment \cite{KOBUSINSKA20181321, ikramSeamlessTrackingFreeWeb2017} entropy-based detection of fingerprinting.

\subsection{Programmatic labeling}
\label{sec:programmatic-labeling}

In supervised and semi-supervised learning an input label set is required in order to train the model. Labeling is typically done manually and can be very time consuming.

In the case of tracking detection, expert reviewers are required to ascertain the intention of scripts, as noted by \cite{ikramSeamlessTrackingFreeWeb2017} who also suggest exploration of semi-supervised approaches due to this limitation.

With our dataset organized into snippets, we now build a set of labeled snippets from which other snippets are labeled and, in turn, scripts are scored.

\subsubsection{Script finding strategies}
\label{sec:script-finding-strategies}

Leveraging our insight that the execution pattern of scripts with similar behavior will be similar, we do not need a perfect label set. 
We propose two techniques to build the seed label set.

\textbf{Leveraging existing lists} We can use existing lists of fingerprinting or tracking scripts. These lists can be generated from the existing methods for tracking detection (outlined in Section~\ref{sec:detection}). 

\textbf{Keyword search} A key challenge in tracking and fingerprinting detection is that there is no ground truth -- that is, no large and up-to-date dataset that documents which scripts on the web are and are not tracking or fingerprinting.

To address this challenge, we wanted to assess a labeling strategy that could start with minimal input. To do this we leverage our second key insight: that software engineers write code in human readable forms. 
And, while many engineers minify or obfuscate their work, many don't. 
As a result, we can leverage the text information contained in JS execution datasets and search for keywords that may be of interest, such as ``tracker'' or ``fingerprint.''

There are challenges with using simple text searches like this. For example, the keyword ``tracking'' will miss shorthand versions such as ``trk''.

Finally, these proposed techniques are not mutually exclusive. Real-world implementations would develop a strategy that best meets specific needs. We evaluate these two strategies separately as it provides more opportunity for comparison and analysis.

\subsubsection{Snippet label pruning}
\label{subsec:scripts-to-snippets}

In our model snippets are individual functions name-spaced by their script's domain and name. Fingerprinting scripts, like any other script, use a range of APIs and browser attributes wrapped into different functions. 
Some of these functions are very generic in nature and may be widely present in non-fingerprinting scripts. For instance, \textit{fingerprintjs2} has a function called \textit{getNavigatorPlatform} shown in Figure~\ref{getNavigatorPlatform} that simply returns the platform the browser is running on. 
Functions like \texttt{getNavigatorPlatform} that are present in fingerprinting scripts but also common to a significant proportion of scripts in our dataset are poor labels for our model. However, we do not want to manually review every function in our dataset. 

\begin{figure}[htbp]
\begin{lstlisting}
var getNavigatorPlatform = function (options) {
  if (navigator.platform) {
    return navigator.platform
  } else {
    return options.NOT_AVAILABLE
  }
}
\end{lstlisting}
\vspace{-0.5cm}
\caption{Code Sample from fingerprintjs2 \cite{vasilyevModernFlexibleBrowser2019a}}
\label{getNavigatorPlatform}
\end{figure}

Having generated an input list of scripts, when we map them to their respective snippets they will likely contain mislabels. To programmatically remove mislabels we implement the following pruning strategy. We first compute the pairwise distance matrix as outline in Section \ref{sec:distance-computation} on the full label set. For a range of distance thresholds we build a distribution of the proportion of the dataset that each label is close to. Based on these distributions a threshold is selected. The label set is then pruned to contain only labels which label less than the threshold proportion of the dataset. 

We found that across all experiments and a wide range of distance thresholds there was no ambiguity in selecting a threshold. We saw characteristic bi-modal distributions with narrow peaks and a large gap between them. This process can be automated and our model implementation contains a default pruning threshold of 0.2 based on our findings.

\subsection{Score calculation}
\label{score-calculation}

We now build our model and calculate a score for each script based on the number of labeled snippets a script is close to in our engineered JS execution space. 

\subsubsection{Distance computation}
\label{sec:distance-computation}

Given our $n$ snippets, each represented as a vector over $s$ symbols, we compute pairwise distances between every snippet and each of the $d$ labeled snippets, resulting in a $n \times d$ distance matrix. 

Due to the size of our data, the $n \times d$ matrix quickly gets very large. For 700,000 snippets and 70,000 labels we must compute 49 billion pairwise distances. To make this a tractable computation a vectorized implementation is necessary. This limited our choice of distance metrics. We evaluated the efficacy of various similarity metrics in providing discriminating power and found the Chebyshev distance performed well while also being computationally efficient. The comparison between distance metrics is described in Appendix~\ref{appendix:distance-metric-delta-comparison}, showing potential performance gains through the use of cosine distance, at the cost of a 10-20$\times$ increase in processing time.

The Chebyshev distance only takes into account the dimension in which the two snippets differ the most. It is useful in sparse high-dimensional datasets like ours, as it tends to ignore slight differences that may be due to noise.

\subsubsection{Distance threshold}
\label{sec:distance_threshold}

Next we find a distance threshold. Snippets closer than the distance threshold to a label are similar enough to be included in the expanded set. Snippets further away--those that are less similar--are not.

In order to select a distance threshold $(d)$, we first compute the scores (see Section~\ref{subsec:scoring}) for all scripts at varying distance thresholds. We then order scripts by their score to achieve a ranked list. As in \cite{pantelcrestanetal2009web,zhangliu2011entity} we then look at recall and precision for the ranked script list against the \textit{input list}. Recall and precision can be examined separately, but we found the F1 score, which combines recall and precision, typically gives results that match our intuition and analysis of the data. Although it can break down at low ranks.

\begin{figure}[htbp]
\centering
\includegraphics[width=0.7\columnwidth]{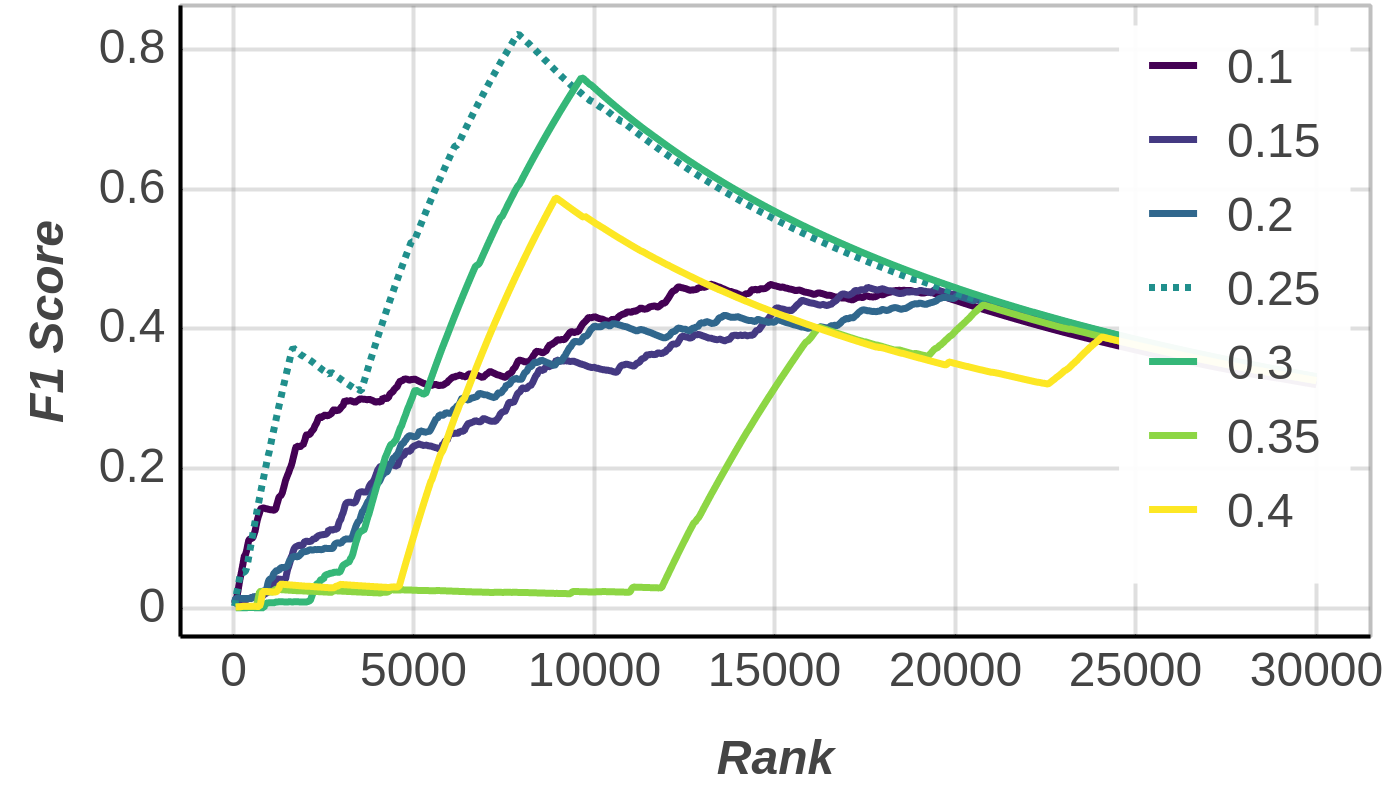}
\caption{Rank F1 Score over varying distance thresholds for threshold distance selection}
\label{fig:distance-threshold-selection}
\end{figure}

Figure~\ref{fig:distance-threshold-selection} shows an example of the Rank-F1 plot from our list labeling experiment. In this plot, the dashed line (d=0.25) is the line of the optimal distance threshold as it achieves the highest F1 score.

\subsubsection{Scoring}
\label{subsec:scoring}
With our collection of $k$ labeled snippets determined, we
restrict attention to the corresponding $n \times k$ subset of the distance matrix.
For each snippet vector $x$, we compute the count $c(x)$ of how many labeled snippets are near it:
\[c(x) = \big|\{y\in D: \delta(x, y) \leq d\}\big|\],

where $d$ is the threshold selected in Section  \ref{sec:distance_threshold}, and $D$ is the the set of labeled snippets.

To compute a final score for a given script, we build a mapping of clean script urls (see Section\ \ref{subsec:get_clean_script}) to snippets.

Each snippet has been assigned the count of how many labeled snippets are near it. The final score for each script is the maximum count observed across the script's snippets. Formally, if $S(r)$ is the set of snippets associated with clean script url $r$, the script score $d(r)$ is:
\[
d(r) = \max_{x \in S(r)} c(x),
\]

\subsection{Validation methodology}
\label{validation-methodology}

We have stressed the lack of currently available ground truth to facilitate tracking and fingerprinting detection. This presents a problem when we wish to validate our methodology. To do this we undertake a process similar to Gugelmann et al. \cite{gugelmannAutomatedApproachComplementing2015} and Iqbal et al. \cite{iqbalAdGraphMachineLearning2018}. First we compare our results with an available reference list. Then we manually evaluate a sample of our remaining flagged scripts to characterize our results. In our case, the available reference list is a list of fingerprinting scripts built using existing heuristics based on OpenWPM datasets outlined in the literature by Englehardt \& Narayanan~\cite{englehardtOnlineTracking1millionsite2016} and Das et al.~\cite{dasWebSixthSense2018}.

To evaluate the remaining results, we made the problem more tractable by first looking for identifiable collections of scripts. 
For example, the modernizr script \cite{ModernizrFeatureDetection} manifests as a distinctive combination of symbols and arguments in the data. 
These collections are outlined in Appendix~\ref{appendix:collections}.

We were then left with a remaining set of ``uncharacterized'' scripts. From these we performed a manual review of a sample of scripts. Two of the present authors first independently assessed 20 randomly selected scripts, and compared results to refine agreement and standards of assessment. Then an additional sample of 103 scripts was randomly selected and independently reviewed for the following criteria: Is the script fingerprinting? If yes, what types of fingerprinting?; is the script engaging in tracking (not necessarily fingerprinting)? and finally Is the script associated with a benign use of the HTML Canvas API? The question about HTML Canvas was motivated by anecdotal observations of our results while developing our methodology. The reviewers achieved 90\% or higher agreement in their script assessment across all criteria.

With manual review complete for scripts above a given score, we can report a false positive rate and the rank-precision or precision$@$rank value as used for evaluation in the entity set expansion literature \cite{zhangliu2011entity,pantelcrestanetal2009web}.

\section{Evaluation}
\label{sec:findings}
This section details the implementation of our model and the OverScripted dataset \cite{1-OverscriptedDataSet2018} used to evaluate it. We perform the two experiments to assess our model's ability to detect fingerprinting scripts. \textbf{Experiment 1} uses an input list obtained by applying the Das et al. \cite{dasWebSixthSense2018} heuristics. \textbf{Experiment 2} uses a keyword-based approach to create an input list.

Our model performs comparably well in both experiments, showing its resilience to the input list.
Finally, we show how our model is able to discover fingerprinting scripts that were not present in the input lists. In particular, we highlight our discovery of device-class fingerprinting in the wild, a new class of fingerprinting surfaced by our model.

\subsection{Dataset description}
\label{dataset-overview}

We apply our methodology to the publicly available OverScripted dataset \cite{1-OverscriptedDataSet2018} collected using OpenWPM \cite{webtapResearch, WebPrivacyMeasurement2019}. Details of the crawl are outlined in \cite{3-lopatkaOverscriptedDiggingJavaScript2019}. 
The dataset consists of \textasciitilde{}114 million JS calls from \textasciitilde{}2.1 million location URLs (locations may be navigated-to pages or locations in nested iframes).
The \textasciitilde{}114 million calls originate from \textasciitilde{}1.3 million unique script URLs, which reduce to \textasciitilde{}407,000 scripts when cleaned of query strings, fragments, and scheme (http/https).

We grouped the data as outlined in Section~\ref{sec:snippet_preparation}, resulting in a dataset with \textasciitilde{}793,000 snippets. The dataset contained 282 unique symbols. \footnote{This is a reflection of the symbols that were instrumented by OverScripted / OpenWPM, and not representative of the breadth of JS APIs that may be called by web pages, of which there are thousands.} The result is a sparse initial matrix 282 x 793,000.

As the OverScripted dataset does not include script contents, we attempted to retrieve the scripts between Nov 2018 and Feb 2019. Most scripts were retrieved. The scripts were used for validation purposes as outlined in Section~\ref{validation-methodology} and for exploring the device-class fingerprinting scripts we discovered as described in Section~\ref{sec:device_class_fp}. In the 22 cases where the sampled scripts were not available for review, we relied on the execution data present in the dataset, 
which includes recognizable patterns.
Additionally, there is a possibility that the script we retrieved for a given URL is different from that retrieved and executed during the crawl. However, we believe that this has not impacted our results as the findings based on the dataset were not in disagreement with any instances of manual inspection.

\subsection{Model implementation}
\label{subsec:model-implementation}

With hundreds of thousands snippets and of tens of thousands of labels, the resulting pairwise distance matrices contains tens of billions of entries. 
To make these computations tractable, xarray \cite{NDLabeledArrays2019} was used to parallelize the computation in both the row and column dimensions. 
The full source code of the model is available at \cite{UtilitiesBuildDyescore2019}.

\subsection{Heuristic list compilation}
\label{heuristic-list}

Studies from Englehardt \& Narayanan \cite{englehardtOnlineTracking1millionsite2016}, Das et al. \cite{dasWebSixthSense2018}, and this paper are all based on OpenWPM for web measurement, so we are able to leverage these earlier studies to build a list of known fingerprinting scripts.

While the limitations of heuristics are motivation for our work, a list based on these published heuristics is the best available option we have to generate a ground truth for our dataset. We use this list in two distinctly different ways: 
\begin{enumerate}
\def\labelenumi{\arabic{enumi}.}
\item
  As a label set as described in Section \ref{sec:script-finding-strategies} and used for Experiment 1 (Section~\ref{sec:exp1-list-labeling}).
\item
  As a reference list to aid in validating our methodology as described in \ref{validation-methodology}.
\end{enumerate}

The heuristics detailed by Englehardt \& Narayanan \cite{englehardtOnlineTracking1millionsite2016} and Das et al. \cite{dasWebSixthSense2018} differ only slightly on a logical level for canvas, canvas font and audio fingerprinting. 
However, small implementation changes result in differences in the flagged scripts. We summarize this in Table \ref{tbl:heuristic_list_stats} using the Jaccard similarity---the size of the intersection of two
sets relative to the size of the union of those sets. 
We found discrepancies between the two implementations, which are detailed in Appendix~\ref{heuristic_implementation}. We chose the list generated from Das et al.'s work because it was more conservative and the code is publicly available \cite{MobileWebPrivacy2019}.

\begin{table}[htbp]
\centering
\begin{tabular}{ccccc}
 & Canvas  & Canvas Font & WebRTC & Audio  \\
\midrule
(a) & 8,503  & 1,387 & 1,313 & 534      \\
(b) & 8,519   & 1,387 & 1,313 & 170      \\
Similarity & 87.00\% & 100\% & 100\% & 31.34\%  \\
\bottomrule
\end{tabular}
\vspace{0.2cm}
\caption{Number of scripts found using heuristics from \\
(a) Englehardt \& Narayanan \cite{englehardtOnlineTracking1millionsite2016} and
(b) Das et  al. \cite{dasWebSixthSense2018}}
\label{tbl:heuristic_list_stats}
\end{table}

These URLs gathered from applying the heuristics are cleaned (see \ref{subsec:get_clean_script}) to produce our final heuristic list of 6,028 scripts with the following breakdown: Canvas (5,947), Canvas Font (52), WebRTC (132) and Audio (59). We note the dramatic drop in script count for Canvas Font and WebRTC highlighting that these fingerprinting techniques are served by just a few providers. For example \texttt{m.stripe.network/inner.html} accounts for 971 of the raw Canvas Font urls.

\subsection{Experiment 1: Labeling with heuristic list}
\label{sec:exp1-list-labeling}

For our first experiment, we compile a label list from the collection of scripts in our heuristic reference list. That is, labeled snippets are from scripts identified by Das et al.'s heuristics \cite{dasWebSixthSense2018} as canvas, canvas font, audio, or webrtc fingerprinting. With a selected distance threshold of 0.25 and label pruning threshold of 0.2, the scripts were scored.

\textbf{Validation - Performance against reference list} As described in Section \ref{validation-methodology} to validate our results we first compare our results with our heuristic reference list that was used as an input. It may seem like a self-fulfilling prophecy to validate against our input list. However, the model operates in snippet-space not script-space and it's important to understand how our programmatic labeling strategy operates moving from scripts to snippets and back again. 

\begin{figure}[htbp]
\centering
\includegraphics[width=0.7\columnwidth]{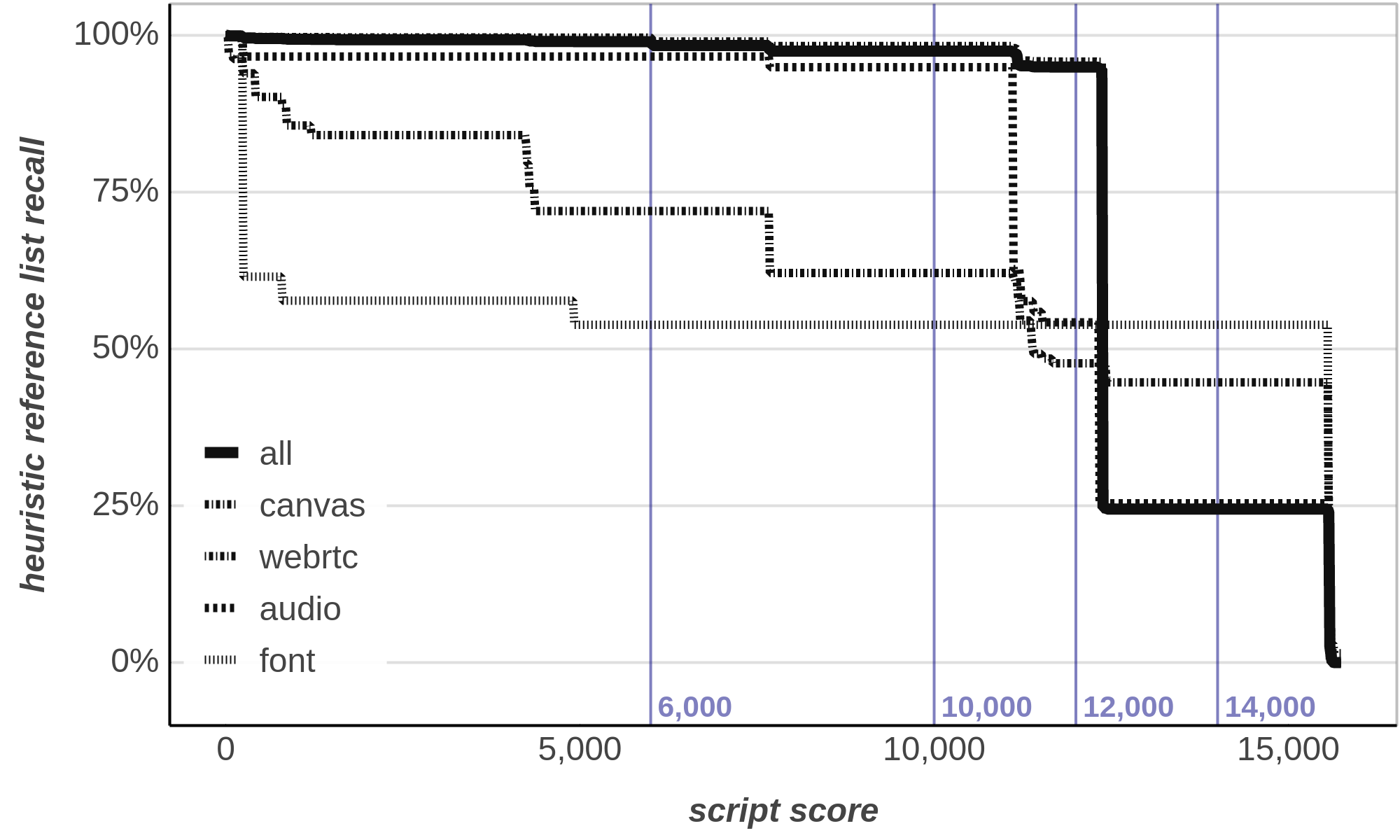}
\caption{Validation Plot - List labeling strategy performance compared to reference heuristic list and its four sub-components (canvas, font, webrtc, audio).}
\label{fig:list_perf}
\end{figure}

In Figure~\ref{fig:list_perf} we examine the recall rate for the heuristic reference list (all) and each of its sub-components: canvas, webrtc, audio, and font. This lets us explore in more detail how the model is performing. We plot recall against model's script score output. 

The score of 12,000 marked in Figure~\ref{fig:list_perf} corresponds to the peak of the Rank-F1 plot shown in Figure~\ref{fig:distance-threshold-selection}. This maxima is only compared to our heuristic list, which we know has limitations, and we therefore review the scripts flagged at four different scores corresponding to plateaus in Figure~\ref{fig:list_perf}. The results summarized in Table \ref{tbl:script_review_summary}.

\begin{table}[htbp]
\centering
\begin{tabular}{llrrrr}
 & Score                        & 6,000   & 10,000 & 12,000          & 14,000 \\
\midrule
 &  n scripts over score        & 28,978 & 18,961 & 7,915 & 3,434  \\
 & \% of dataset       & 7.1\% & 4.6\% & 1.9\% & 0.8\% \\
\midrule
\multicolumn{5}{l}{\textbf{Pre-characterize}} \\
 & n in heuristic list                 & 5,962  & 5,875  & 5,722           & 1,476  \\
 & \% of heuristic list                & 98.9\% & 97.5\% & 94.9\% & 24.5\% \\
\multicolumn{5}{l}{fingerprinting} \\
 & akam                         & 108     & 108     & 108       & 0     \\
 & hs                & 1,857    & 1,857    & 0      & 0 \\
\multicolumn{5}{l}{benign use of canvas} \\
 & charting                    & 151     & 104     & 87        & 87     \\
 & modernizr                    & 2,588     & 453     & 453       & 453    \\
 & sadbundle                    & 335     & 281     & 258       & 258     \\
\midrule
\multicolumn{5}{l}{\textbf{Remaining uncharacterized}} \\
 &                          & 17,977 & 10,283  & \textbf{1,287}    & 1,160    \\ 
\bottomrule
\end{tabular}
\vspace{0.2cm}
\caption{Validation summary Summary of scripts grouped at selected score thresholds. Uncharacterized scripts from the 12,000 score group were manually reviewed.}
\label{tbl:script_review_summary}
\end{table}

Of the 7,915 scripts given a score of 12,000 or higher by the model, 5,722 (94.9\%) are fingerprinting as identified by the heuristic list. This increases to 5,962 (98.9\%) at a score of 6,000.

As we look at the sub lists, we see that canvas fingerprinting scripts follow the overall line closely. 
The audio and webrtc follow a steady progression of increasing recall as the score is lowered. 
For canvas font fingerprinting, however, we see a noticeably different result we readily capture \textasciitilde{}$50$\% of the known font fingerprinting scripts, and then struggle to improve on that. 
We discuss this further in Section \ref{discussion}.

Having looked at our reference list, we can review the performance of the model to see what we've
scored highly that's not already known as fingerprinting via heuristics.

\textbf{Validation - Script collections} We identified collections of scripts that were not labeled by the heuristics but that we could manually identify as distinct sets of scripts. The collections were named ``akam'', ``hs'', ``charting'', ``modernizr'', and ``sadbundle''. They are discussed in detail in Appendix~\ref{appendix:collections}. In brief, ``akam'' and ``hs'' appear to be fingerprinting scripts. The others are legitimate uses of the HTML Canvas API.  

At a score of 12,000 108 ``akam'' fingerprinting scripts are flagged along with 798 from our collections that are legitimate uses of the HTML Canvas API. At a score of 10,000 we capture a large block of ``hs'' attribute fingerprinting scripts.  It's worth emphasizing that the model is giving high scores to browser attribute fingerprinting scripts for which there are currently no heuristics.

\textbf{Validation - Manual review of uncharacterized} Due to the labor-intensive nature of manual review, we only review at the score of 12,000. We manually reviewed a sample of 103 scripts out of the 1,287 uncharacterized scripts. Of the sample, we found 22 (21\%) to be fingerprinting. These were a range of types of fingerprinting. There were a few instances of \texttt{fingerprint2.js} that would typically be picked up by heuristics, but were not due to some APIs not being called. This could happen because a page is slow to load during the crawl and offers an example of how the model can be robust to measurement fluctuations. We also detected a number of attribute fingerprinting scripts for which there are no heuristics. Finally, our sample contained 7 scripts that appeared to be device-class fingerprinting. This is covered in more detail in Section~\ref{sec:device_class_fp}.  

The remaining scripts in the sample were false positives. 67 scripts (65\%) were benign uses of the HTML Canvas API ranging from media players to small canvas widgets along with many more examples of modernizr \cite{ModernizrFeatureDetection} that were not included in our conservative automatic classification. 14 final scripts (14\%) were neither fingerprinting nor benign uses of the HTML Canvas API and we marked as other. They had a range of functions including ad-tech, tracking scripts, and JS applications.

\textbf{Validation - Summary} Collecting together our results from the steps above we provide the following summary.

The total number of scripts given a score of 12,000 or higher was 7,915. Of these, the total number of fingerprinting scripts was 6,100 (77\%) with 378 of these being new detections not covered by the heuristic reference list.

1,635 (21\%) scripts that were a benign use of the HTML Canvas API and 180 (2\%) other scripts were also given a score of 12,000 or higher for a false positive rate at a script score of 12,000 of 23\%.

Having performed a manual review and established an estimated ground truth within our sample we can report our model's precision of 77\% at a score of 12,000, or as a rank-precision$@$7915. As we do not know the true fingerprinting prevalence in our dataset, we cannot report overall recall. We note that these results are only valid at a score of 12,000 and cannot be projected to the model's output at different script scores. We can see from the script collections that at a score of 10,000 a large swath of browser attribute fingerprinting scripts get flagged while flagging of benign use of canvas increases only fractionally. It may be that the ``uncharacterized'' scripts at this level are disproportionately browser attribute fingerprinting scripts. 

\subsection{Experiment 2: Keyword-based labeling}
\label{sec:keyword-labeling}

In Experiment 1 we started with known fingerprinting scripts and found other scripts that exhibited similar execution patterns. In this experiment, we built our label set by searching for a keyword in the \texttt{script\_url} and \texttt{func\_name} fields. 
We looked at a number of keywords and found, perhaps unsurprisingly, that ``fingerprint'' gave the best results.

The ``fingerprint'' keyword was found in 462 scripts (3k snippets). This is a small fraction of the 6k scripts (70k snippets) generated from our heuristic reference list. With a selected distance threshold of 0.25 and label pruning threshold of 0.2, the scripts were scored. 

\begin{figure}[htbp]
\centering
\includegraphics[width=0.7\columnwidth]{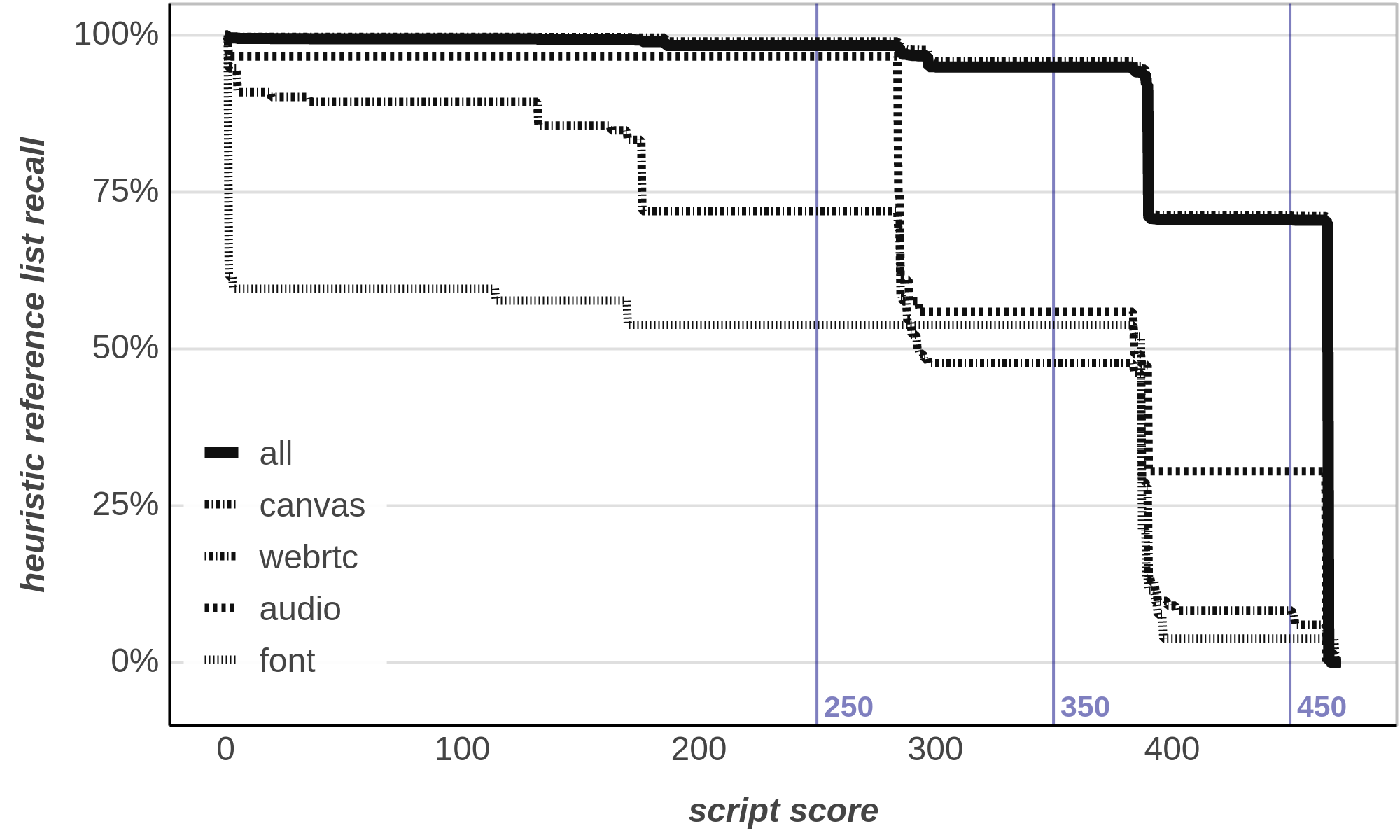}
\caption{Validation Plot - Keyword "fingerprint" labeling performance compared to reference heuristic list and its four sub-components (canvas, font, webrtc, audio).}
\label{fig:kwdyeing_perf}
\end{figure}

\textbf{Validation} In Figure~\ref{fig:kwdyeing_perf} we show the performance of the ``fingerprint'' keyword input list with recall curves for our heuristic reference lists. 
The lower scores are an artifact of the smaller input list.
The profile is very similar to list labeling in Experiment 1. This is a surprising and positive result. Simply searching for ``fingerprint'' in our dataset, and using that as the basis for labeling scripts based on pairwise distances of snippets in our vector space, we are able to identify 94.9\% of the scripts identified by heuristic-based methods.

Table \ref{tbl:keyword_summary} summarizes the performance as we did for Experiment 1. At the 350 score threshold we get very similar performance to that seen in Experiment 1 at a score of 12,000: we get essentially the same recall rate compared to the heuristic-based reference list and we picked up the same set of device-class fingerprinting scripts. An improvement over Experiment 1 is that we do \textit{not} capture a large chunk of modernizr scripts. The uncharacterized scripts from list labeling score of 12,000 and keyword labeling score of 350 have a Jaccard similarity of 75\%. No obvious patterns jumped out as to the scripts that were captured by one and not the other (n=85). 

\begin{table}[htbp]
\centering
\begin{tabular}{llrrr}
 & Score                        & 250   & 350 & 450 \\
\midrule
 &  n scripts over score        & 28,927 & 7,305 & 4,491  \\
 & \% of dataset       & 7.1\% & 1.8\% & 1.1\% \\
\midrule
\multicolumn{5}{l}{\textbf{Pre-characterize}} \\
 & n in heuristic list                 & 5,929  & 5,723           & 4,255 \\
 & \% of heuristic list                & 98.4\% & 94.9\% & 70.6\% \\
\multicolumn{5}{l}{fingerprinting} \\
 & akam                         & 108     & 108     & 108      \\
 & hs                           & 1,857    & 0      & 0 \\
\multicolumn{5}{l}{benign use of canvas} \\
 & charting                    & 151    & 89        & 0     \\
 & modernizr                   & 2,588     & 0     & 0   \\
 & sadbundle                   & 335     & 270           & 0     \\
\midrule
\multicolumn{5}{l}{\textbf{Remaining uncharacterized}} \\
 &                          & 17,959  & 1,115    & 128    \\ 
\bottomrule
\end{tabular}
\vspace{0.2cm}
\caption{Validation Table - Summary of scripts flagged by keyword ``fingerprint'' grouped at selected score thresholds.}
\label{tbl:keyword_summary}
\end{table}

\subsection{Detecting device-class fingerprinting}
\label{sec:device_class_fp}
Our method creates a candidate set of scripts that are likely to contain fingerprinting. This allows us to filter through the noise of the \textasciitilde{}400k scripts contained in our dataset to a much smaller set that can be manually reviewed. After filtering out classes of known scripts flagged by our model in Experiment 1 (Section~\ref{sec:exp1-list-labeling}), we were left with only 1,287 uncharacterized scripts. A manual review of a small sample uncovered 7 scripts that appear to participate in device-class fingerprinting using 2 different methodologies. To the best of the authors' knowledge, this is the first time device-class fingerprinting scripts have been documented in the fingerprinting measurement literature. This shows how our model allows us to---with relatively little manual effort---expand beyond the constraints of our labeled input data and discover new techniques.

\subsubsection{Device-class fingerprinting using canvas} 
\label{sec:device_class_fp_using_canvas}

We discovered a first-party Facebook script that appears to be a custom implementation of the Picasso device-class fingerprinting method \cite{burszteinPicassoLightweightDevice2016}. Instances of the script are always served from \texttt{*.facebook.com} and are present on pages across Facebook's website, but are not served from paths with a discernible pattern. The script includes code to append the result of the challenge to a hidden element in a login form. We posit that the data could be used to mitigate attacks on Facebook's authentication infrastructure.

\begin{figure}[htbp]
\captionsetup[subfigure]{justification=centering}
\centering
\begin{subfigure}{.4\linewidth}
\centering
\fbox{\includegraphics[scale=0.44]{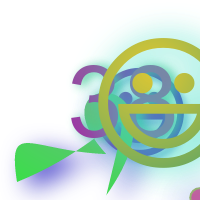}}
\caption{Mixed primitives}\label{fig:picasso_mixed}
\end{subfigure}
\begin{subfigure}{.4\linewidth}
\centering
\fbox{\includegraphics[scale=0.44]{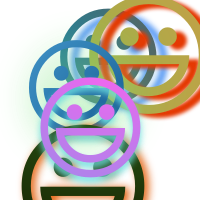}}
\caption{Emoji primitive}\label{fig:picasso_smileys}
\end{subfigure}
\\
\begin{subfigure}{.4\linewidth}
\centering
\fbox{\includegraphics[scale=0.44]{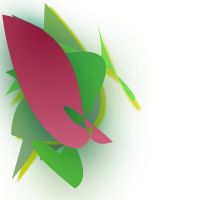}}
\caption{B\'ezier primitive}\label{fig:picasso_curves}
\end{subfigure}
\begin{subfigure}{.4\linewidth}
\centering
\fbox{\includegraphics[scale=0.44]{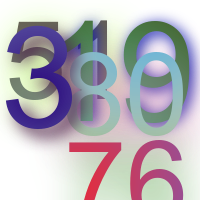}}
\caption{Number primitive}\label{fig:picasso_numbers}
\end{subfigure}
\setlength{\abovecaptionskip}{10pt}
\caption{Four example canvases from the Facebook script which appears to implement Picasso-style \cite{burszteinPicassoLightweightDevice2016} device-class fingerprinting. We include a mixed primitive example  (a), as well as several canvases that result from multiple iterations of the same primitive (b-d).}\label{fig:facebook-picasso-example}
\end{figure}

The Facebook script follows a very similar design to Picasso, but differs in the types of graphical primitives it implements. Picasso generates a device-class fingerprint by drawing a randomized graphical primitive to canvas over multiple rounds. In each round, the Picasso script will select a primitive, choose a random color and shadow to apply, and hash the result. The Facebook script implements four primitives: a \textit{textual} primitive that writes a random integer between 0 and 100 to a random location on the canvas, an \textit{emoji} primitive that writes the \includegraphics[scale=0.6]{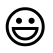} emoji character (unicode character \texttt{U+1F603}) to a random location on the canvas, a \textit{circle} primitive that uses the \texttt{arc} method of the rendering context to draw a circle, and a \textit{bezier} primitive that draws several B\'ezier curves. For all four primitives, a randomized style and shadow is applied. Like Picasso, the script appears to support running different combinations of these primitives in multiple rounds and computes the hash of the resulting canvases. In Figure~\ref{fig:facebook-picasso-example} we show an example canvas image drawn during one execution of the script.

While these operations are conceptually very similar to the types of canvas operations expected from a canvas fingerprinting script, these scripts are missed by the heuristics implemented by Das et al. \cite{dasWebSixthSense2018} and Englehardt \& Narayanan \cite{englehardtOnlineTracking1millionsite2016}. The Facebook scripts writes only a short integer string (e.g., "23") or a single-character emoji. The string length and complexity requirements of both Das and Englehardt's work will filter out these scripts.

Both labeling experiments identified all the scripts we believe are in the dataset that perform this type of fingerprinting.

\subsubsection{Device-class fingerprinting using feature inspection} 
\label{sec:device_class_using_feature_inspection}

We discovered a second collection of scripts that appear to perform device-class fingerprinting through deep feature inspection. In addition to inspecting for the existence of certain APIs, these scripts appear to search for implementation differences in certain JS properties. This is in contrast to a typical feature detection script (e.g., Modernizr \cite{ModernizrFeatureDetection}), which attempts to discover the set of available APIs that the device supports. For example, we observe the scripts inspect the return types for several APIs (e.g., \texttt{"boolean" == typeof a.navigator.onLine} and \texttt{"number" == typeof a.performance.now()}) and examine the initialization values for new arrays (e.g., \texttt{0 == (new Uint16Array(1))[0]}).

We also observe compound tests, where the script inspects multiple unrelated APIs. In Figure~\ref{fig:canvas-vibrate-inspection} we include a simplified code snippet of one such example. In this snippet, we observe the script create a 1 pixel by 1 pixel canvas, set the canvas to multiply the RGB values of overlapping pixels, and then draw two overlapping rectangles of different colors. Finally, the script checks whether certain channels of the resulting pixel are equal. If this check fails, the script instead checks whether the \texttt{window.navigator.vibrate} API exists. Given that these two checks are entirely unrelated, we posit that the return values of this compound series of checks is different across different devices.

\begin{figure}[htbp]
\begin{lstlisting}
function() {
  canvas = document.createElement("canvas");
  canvas.width = canvas.height = 1;
  ctx = canvas.getContext("2d");
  ctx.globalCompositeOperation = "multiply";
  ctx.fillStyle = "rgb(0, 255, 255)";
  ctx.fillRect(0, 0, 1, 1);
  ctx.fill();
  ctx.fillStyle = "rgb(255, 255, 0)";
  ctx.fillRect(0, 0, 1, 1);
  ctx.fill();
  d = ctx.getImageData(0, 0, 1, 1).data;
  return (d[0] == d[2] && d[1] == d[3])
    || exists(window.navigator.vibrate);
}
\end{lstlisting}
\vspace{0.1cm}
\caption{A script that inspects both the Canvas API and the Vibration API as part of a device-class fingerprint. The result of inspecting these unrelated APIs is combined into a single feature. This is a simplified version of a function found in \url{imasdk.googleapis.com/js/core/bridge3.166.0_en.html}.}
\label{fig:canvas-vibrate-inspection}
\end{figure}

The model flagged a majority of the variants of this type of device-class fingerprinting at our chosen thresholds. We determined prevalence by searching in our corpus of scripts downloaded for manual review for the characteristic function shown in Figure~\ref{fig:canvas-vibrate-inspection}. We consider this a lower bound, as many scripts could not be retrieved. The detection methodology is outlined in Appendix~\ref{appendix:collections}. We found 121 unique ``clean'' scripts served from 13 distinct eTLD+1 domains (listed in Appendix~\ref{dcfp}). These were embedded on 608 unique eTLD+1 domains visited during the crawl.

We report score results from experiment 1 (list labeling) for the device class scripts. A script score of over 15k (v. high) was given to 30 of the 121 scripts covering 89\% (542) of the eTLD+1 domains where the script was found. A score in the 4,000s was given to a further 63 scripts covering 81 domains, and finally a low score was given to the remaining 28 scripts present on 14 eTLD+1 domains.

Similar to the Facebook script examined above, the canvas heuristics developed by Das et al. \cite{dasWebSixthSense2018} and Englehardt \& Narayanan \cite{englehardtOnlineTracking1millionsite2016} fail to capture these scripts. The type of feature detection carried out does not draw large canvases and does not write any text to the canvas -- which both Das and Englehardt's detection heuristics require. This further highlights the strength of our technique in discovering related fingerprinting techniques (i.e., fingerprinting for device-class discovery rather than device identification) and fingerprinting carried out using techniques that are harder to build heuristics for such as attribute fingerprinting. We did not set out looking for device-class fingerprinting, but the model results prompted our discovery.

\section{Discussion}
\label{discussion}

Our findings reveal three themes with which we can frame the current capabilities and limitations of the model and suggest avenues for future work: deepening precision, extending the methodology, and complementing existing techniques. 

\textbf{Deeper precision}.
Despite our ability to detect a wide range of fingerprinting scripts, including instances of attribute and device-class fingerprinting that are not detected by the state-of-the-art heuristics, we note two key shortcomings in our model. 
First, we found legitimate uses of the HTML Canvas API that we incorrectly flagged. Blocking these scripts may lead to site breakage.
The may be acceptable to advanced privacy tool users but would be hard for providers of privacy enhancing tools seeking more general appeal to promote.

Second, we failed to flag nearly 50\% of known font fingerprinting scripts.
We believe this is because our vector representation lacks a feature specific to this type of fingerprinting, namely repetitive access to a small number of APIs, the scripts we missed being those that only do font fingerprinting.

Additionally, we would like to improve the granularity of our snippet labeling, allowing a flexible labeling of only those functions that are responsible for fingerprinting.
And finally, fine-grained tuning would allow users to adjust their inputs to the model as well as the desired degree of similarity in order to build lists that suit their use case. This would enable list maintainers to leverage the scalability of the model while controlling the confidence they need in the resulting output, in essence, creating a configurable privacy threshold.

\textbf{Extending the model}. Our methodology has core features that make various extensions viable. 
Firstly, experimentation should be done in varying input lists, including adding fuzzy searches to keyword labeling, \textit{e.g.}, to catch `traking` along with `tracking`.
Another extension would be to add negative labels. That is, providing input lists of items that we know are safe and should not be labeled as fingerprinting.

The model is flexible and allows combining input strategies (\textit{e.g.}, keyword labeling, heuristics), or feeding the output of one round of labeling to the next. This could be done on a single dataset to integrate the results of multiple inputs, or it could be used across longitudinal datasets to examine drift or trends.

In our methodology, we group JS calls into snippets but we do not give importance to the order of these calls.
The order of calls could be used to create a \textit{function signature} that would be more reliable and harder to evade than using function names.
Because we believe that many of the fingerprinting functions are reused, shared and copied into many different scripts, we argue the use of function signatures would improve the detection of such scripts and allow for detecting modified scripts that perform, in essence, the same operations.
This should improve our model's resistance to deliberate obfuscation or optimizations introduced by transpilers.

Furthermore, the current dataset we use instrumented 282 APIs that are suspected to be used for fingerprinting. This has proven effective in correctly labeling fingerprinting scripts, but using such a targeted set of APIs leads to benign scripts being incorrectly labeled.
We believe that increasing the number of APIs would benefit our model by filtering out benign scripts because they use a broader range of APIs and fingerprinting scripts tend to be characteristically narrow in their API selection.
Given the cost of an exhaustive instrumentation of APIs, one approach to further increase the set of instrumented APIs in a targeted manner may come from JS templating attacks \cite{javascriptTemplateSchwarz2017}.

\textbf{Complementing existing methodologies}. We are optimistic about opportunities to combine our model with other techniques for fingerprinting detection. 
We defined fingerprinting as the use of browser attributes to generate an identifier. 
However, both heuristics and our model target the accessing of browser attributes, not ID generation. 
Work by the Cliqz group \cite{macbethTrackingTrackersAnalysing2016,karajWhoTracksMeMonitoring2018} looked for unique ID transmission.
We believe combining this technique with our model would improve our results. 
In particular benign uses of the canvas API are unlikely to be generating unique identifiers. 
However, as the Cliqz methodology in its published state uses actual user browsing data, it was out of scope to consider its inclusion for this paper.

\textbf{Limitations} Our work is based on crawl data. There is no guarantee that the crawl data reflects a real users' exposure to fingerprinting. 
For example, fingerprinting may only occur when a user interacts with certain parts of a page (e.g., attempts to register for an account), or after a certain time delay that's beyond our crawl configuration.
Should the crawler be detected as such, pages may respond with different content than to organically generated traffic. 
Nevertheless, we do not expect these potential differences to impact the validity or evaluation of our proposed model.

Our work assumes that fingerprinting scripts do not actively split their content across multiple files. While the lack of any large-scale deployment of fingerprinting script detection and blocking (Section~\ref{sec:anti-fingerprinting-techniques}) means this is likely a safe assumption, future blocking by browsers could lead to an arms race that encourages such circumventions. To combat this, our method could aggregate script calls at a broader level (e.g., at the page or frame level). An analysis of such circumventions is left to future work.

\section*{Conclusion}
\label{sec:conclusion}
Browser fingerprinting is the creation of unique identifiers composed of attributes collected from the user's device. Fingerprinting is invisible to users and does not require device storage. Furthermore, fingerprinting scripts use APIs that are commonly used in non-fingerprinting scripts, making them difficult to differentiate. In this paper we propose a novel semi-supervised methodology that uses lists of known fingerprinting scripts to detect unknown fingerprinting scripts.

Our key insight is that fingerprinting scripts exhibit similar behavior in generating fingerprints. We present a representation of these API features that is measurable at the per-script (as opposed to per page) level. This vector representation of JS calls made by scripts, which we call snippets, is then used to perform pairwise comparisons between scripts. Snippets known to engage in fingerprinting serve to identify similar behavior in unknown fingerprinting scripts, providing a scalable solution for label propagation when identifying fingerprinting scripts.

Although an input list of known fingerprinting scripts is required, we have shown two strategies for their construction. The first, using heuristics from the state-of-the-art, and for the second, we proposed a simple yet effective solution using keywords.
Our model is more flexible than current state-of-the-art heuristics and was able to detect scripts for which there are currently no detection heuristics, as well as a new class of fingerprinting in the wild, namely, device-class fingerprinting.

We believe that our work also opens a new avenue for automated list-generation and refinement of list-based fingerprinting detection strategies.

\section{Acknowledgements}
\label{acknowledgements}
\addcontentsline{toc}{section}{Acknowledgements}
Special thanks to Jason Thomas and operations team at Mozilla for assistance running crawls. Thanks to all members of UCOSP cohorts that worked on the OverScripted dataset.

\bibliographystyle{ieeetr}
\bibliography{Bibliography.bib}

\appendix
\section*{Appendices}
\addcontentsline{toc}{section}{Appendices}
\renewcommand{\thesubsection}{\Alph{subsection}}

\subsection{Heuristic implementation}
\label{heuristic_implementation}

In Section~\ref{heuristic-list} we discuss that different implementations
of heuristics lead to different results. The results are summarized in Table \ref{tbl:heuristic_list_stats}. In this section we describe the differences found in detail. We assume a familiarity with the heuristics outlined in the two papers and implemented in code released alongside.
Heuristics were defined by examining the code from both Englehardt \& Narayanan \cite{englehardtOnlineTracking1millionsite2016} and Das et al. \cite{dasWebSixthSense2018}. 

The heuristics for canvas font and WebRTC fingerprinting from the two papers produced the exact same set of scripts. That is, the sets had a Jaccard similarity coefficient of 100\%. 
However, for canvas fingerprinting the Jaccard similarity coefficient was 87.00\% and for audio fingerprinting it was 31.34\%. 

\textbf{Canvas} The heuristics used for canvas fingerprinting resulted in two sets of
scripts with a similarity of 87.00\% and 8,503 scripts found 
for \cite{englehardtOnlineTracking1millionsite2016} vs 8,519 for
\cite{dasWebSixthSense2018}. This is due to 3 differences in
implementation. The main difference is that
\cite{englehardtOnlineTracking1millionsite2016} does not explicitly
handle non-ASCII characters while \cite{dasWebSixthSense2018} does
since Python miscalculates unicode strings with characters such as
emojis. Therefore, \cite{dasWebSixthSense2018} drops all non-ASCII
characters. Additionally, in
\cite{englehardtOnlineTracking1millionsite2016}, the core heuristic
was whether a script wrote 10 or more characters \textbf{or} whether it
used 2 or more styles. However,
\cite{englehardtOnlineTracking1millionsite2016} viewed this as an
\textbf{and}. Since \cite{dasWebSixthSense2018} was able to find 
scripts that carried out fingerprinting but only used one 1 style to 
write to canvas, they dropped that requirement and only evaluated scripts 
that wrote at least 10 characters. Finally, \cite{dasWebSixthSense2018} 
does not check that the canvas element's height and width must be greater 
than 16px. The combination of these changes results in 1,184 scripts being
flagged by only one of the two sets of heuristics.

\textbf{Canvas font} For canvas font fingerprinting the two methods
differed slightly but produced the exact same set of scripts. The only
difference was in how they extracted font information from the call to
\texttt{CanvasRenderingContext2D.font}:
\cite{englehardtOnlineTracking1millionsite2016} used a regex to pull
out the font family from the \texttt{value} column while
\cite{dasWebSixthSense2018} matched on the exact text. Both then
checked for 50 or more calls with distinct values. This change resulted
in one more script being found by \cite{dasWebSixthSense2018};
however, that script was dropped in the second step of the heuristic.

\textbf{WebRTC} For WebRTC fingerprinting, the heuristics were the exact same and so
produced the same set of scripts.

\textbf{Audio} Finally, there was a major difference for audio fingerprinting which
resulted in a similarity between the two sets of scripts found of only
31.34\%. In this case, the heuristics from
\cite{dasWebSixthSense2018} seem to have the benefit of audio
fingerprinting being better understood. The original heuristic developed
in \cite{englehardtOnlineTracking1millionsite2016} only checked if
\texttt{OscillatorNode.start} was called.
\cite{dasWebSixthSense2018} updates that to require a series of five
different calls, a more restrictive metric. As a result, the heuristic
derived from \cite{englehardtOnlineTracking1millionsite2016} finds
534 scripts and that from \cite{dasWebSixthSense2018} finds 170, of 
which only 2 were not found in the set of heuristics from 
\cite{englehardtOnlineTracking1millionsite2016}.

\subsection{Distance metric comparison}
\label{appendix:distance-metric-delta-comparison}

We devised a way to assess the performance of different metrics for our dataset despite having no ground truth positive and negative label sets.

For a positive label set we can use our labels generated from list labeling. For our negative label set we take a sample from all remaining snippets. Although this sample is likely to contain positive labels, the proportion of positive labels should be modest as fingerprinting scripts are only a fraction, say up to 10\%, of our script pool.

We wish to find a distance metric that discriminates between different classes of scripts. That is, we same class scripts (positive-positive or negative-negative) should be similar to each other and different class scripts (positive-negative) should be far from each other.

\begin{figure}[htp]
\centering
\includegraphics[width=\columnwidth]{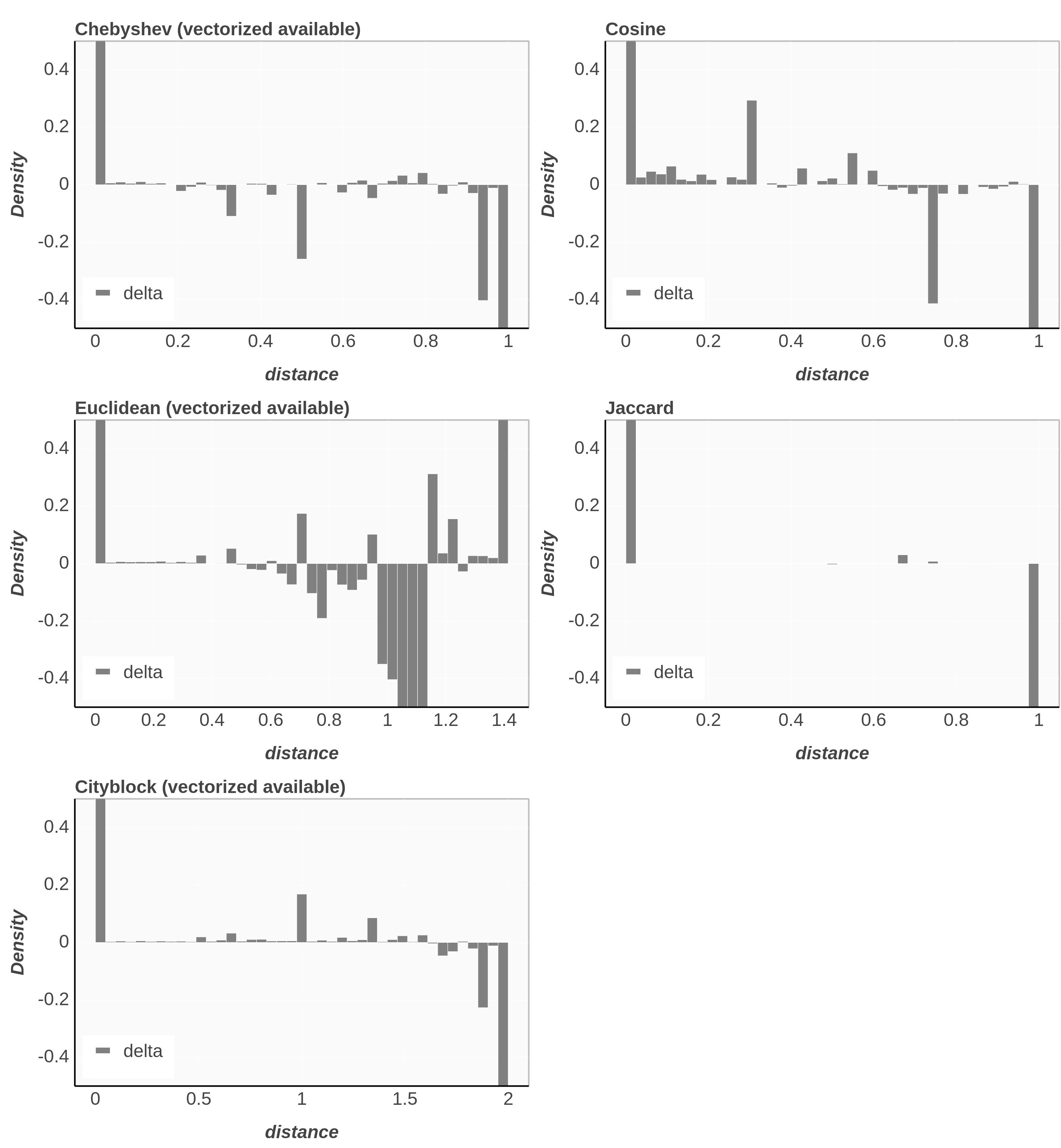}
\caption{Delta between same and different set pairwise distances for 5 distance metrics. Chebyshev, Euclidean, and Cityblock are available in vectorized implementations, Cosine and Jaccard are not meaning that they run 10-20 times slower in current model infrastructure.}
\label{fig:distance-delta}
\end{figure}

We first compute the pairwise distances for all same class scripts and all different class scripts to themselves. We build the distribution of pairwise distances for each class. We then compute the delta - the difference in pairwise distance frequency at a given distance for each different distance metric. If same-class distances dominate at low distances and diff class distances dominate at high distances, the delta will positive at low distances and negative at high distances. This is not a complete evaluation of the relative performance metrics for our complete methodology as the distribution of labels in snippet space matter. However, it gives us some indications. The delta plots are available in Figure \ref{fig:distance-delta}.

We can see that the Chebyshev distance performs well at low distances which is inline with the success of our model as presented in this paper. We can also see taht the Cosine distance appears to offer improvements in discrimination. However, we do not yet have a computationally efficient implementation of cosine and leave this exploration for future work.

\subsection{Script collections}
\label{appendix:collections}

We pulled out identifiable classes of scripts to assist with characterization of the dataset. This section details those classes, and the methods used for identification.

\textbf{akam} The ``akam'' group is built by searching for the string \texttt{/akam/} in the script URL. Although crude, the symbol and argument call values are very consistent across these scripts. Examples of script URLs
are \texttt{https://www.rakuten.co.jp/akam/10/1869811c}, \texttt{https://world.hyatt.com/akam/10/5ea9ce13}, and \texttt{https://www.adidas.ca/akam/10/1b80c866}. 
These scripts were not retrievable with the original recorded script\_url, but going to the locations referenced in the dataset, a script with a similar patterned URL albeit a different suffix was available. This script is obfuscated, but deobfuscation reveals the same script each time. The script contains attribute fingerprinting code along with code for canvas fingerprinting and is blocked, for example, by the EasyPrivacy list \cite{EasyListFilterSubscription2019}.

\textbf{hs} The ``hs'' group is built by searching for the string \texttt{hs-analytics} in the script URL. The symbol and argument calls are consistent, as are the retrieved versions of the script which include a function called \texttt{Fingerprint} which gathers a series of browser properties into an array and hashes the result. 

\textbf{modernizr} The ``modernizr'' group is built by looking at the values set by the script in our dataset. The modernizr script \cite{ModernizrFeatureDetection} has a characteristic set of distinctive values that it often sets. 
\begin{itemize}
    \item 
        The \texttt{userAgent} which for the crawl was \texttt{Mozilla/5.0 (X11; Linux x86\_64; rv:52.0) Gecko/20100101 Firefox/52.0}
    \item
    The dictionary \texttt{\{"modernizr":"modernizr"\}}
    \item
    The empty dictionary \texttt{\{\}}, and
    \item
    The empty string \texttt{""}
\end{itemize}

The modernizr scripts exhibit a variety of other functionalities. In order to be conservative, we only classed scripts that had exactly the set of values described above. This meant there were many modernizr-like scripts in our script review, but we could feel confident we hadn't mis-classed a script that was perhaps bundled with modernizr into this set.

\textbf{sadbundle} The unfortunately named ``sadbundle'' was built by searching for \texttt{tpc.googlesyndication.com/sadbundle/} in the script URL. Existing work with the dataset had revealed that these scripts were always small, unminified, unobfuscated pieces of JavaScript that perform little canvas functions. For example, the URL  \url{http://tpc.googlesyndication.com/sadbundle/$csp\%3Der3\%26dns\%3Doff$/1063502274429671459/JavaScripts/Logo_L.js} which draws a logo or \url{http://tpc.googlesyndication.com/sadbundle/$csp\%3Der3\%26dns\%3Doff$/11515262569827999545/JavaScripts/Ship.js}, which appears to draw a ship on fire!

\textbf{charting} This class was determined after extensive exploration of the dataset. We extracted all scripts which contained ``chart'' or ``jqplot'' and manually reviewed the results. These are all charting or plotting libraries that use the canvas api to render their graphics. The symbol, arguments, and values are all consistent with this activity.

\textbf{device-class fingerprinting} Finally, we describe the detection of the second type of device-class fingerprinting that is characterized by the function listed in Figure~\ref{fig:canvas-vibrate-inspection}. Having built a corpus of scripts by attempting to download every script listed in the datset, we constructed a shortlist by using the unix command \texttt{grep} to find all scripts that contained the word \texttt{vibrate} and the word \texttt{globalCompositeOperation}. From this shortlist of downloaded scripts, we used regular expressions to locate \texttt{vibrate} in the script, then if \texttt{globalCompositeOperation} was in the preceding 500 characters we marked it as belonging to this class of device-class fingerprinting. Manual review of a sample of the ~18,200 scripts returned revealed the expected pattern in all cases examined. A typical sample is: \texttt{ c=c.getContext("2d"); \textbf{c.globalCompositeOperation="multiply"}; c.fillStyle="rgb(0,255,255)"; c.fillRect(0,0,1,1); c.fill(); c.fillStyle="rgb(255,255,0)"; c.fillRect(0,0,1,1); c=c.getImageData(0,0,1,1).data; return c[0]==c[2]\&\&c[1]==c[3]||b.a.bn(\textbf{window.navigator.vibrate}}

\subsection{Device-class fingerprinting}
\label{dcfp}

%The thirteen device-class fingerprinting domains discovered were: 
Scripts served from these domains include what appears to be device-class fingerprinting code:

{'2mdn.net',
 'admaru.com',
 'fqsecure.com',
 'fqtag.com',
 'ibtimes.co.uk',
 'imasdk.googleapis.com',
 'moatads.com',
 'mymovies.it',
 'newsweek.com',
 'securepaths.com',
 'yimg.com',
 'youtube.com',
 'ytimg.com'}

\end{document}